\documentclass[aps,onecolumn,10pt]{revtex4}
\usepackage{amsmath}
\usepackage{amssymb}
\usepackage{hyperref}
\hypersetup{colorlinks=true,linkcolor=red,citecolor=cyan,urlcolor=magenta}
\numberwithin{equation}{section}
\numberwithin{equation}{section}

\begin{document}
\allowdisplaybreaks
\setcounter{equation}{0}

\title{States that grow linearly in time, exceptional points, and zero norm states  in the simple harmonic oscillator}

\author{Philip D. Mannheim}
\affiliation{Department of Physics, University of Connecticut, Storrs, CT 06269, USA \\
philip.mannheim@uconn.edu\\ }

\date{February 19 2026}

\begin{abstract}
The simple harmonic oscillator has a well-known normalizable, positive energy,  bound state spectrum. We show that degenerate with each such positive energy eigenvalue there is a  non-normalizable positive energy eigenstate whose eigenfunction is orthogonal to that of the standard energy eigenfunction. This class of states is not built on the vacuum that $a$ annihilates, but is instead built on the vacuum that $a^{\dagger} a$ annihilates. These non-normalizable but nonetheless stationary energy eigenstates  are accompanied by yet another  set of  non-normalizable states, states  whose wave functions however are not stationary but instead grow linearly in time. With these states not being energy eigenstates, the eigenbasis of the Hamiltonian is incomplete; with the full Hilbert space  containing states that are not energy eigenstates. Thus each energy eigenvalue of the harmonic oscillator is an exceptional point at which  the Hamiltonian becomes of non-diagonalizable, and thus manifestly non-Hermitian,  Jordan-block form. Such non-Hermitian structures  occur for  Hamiltonians that have an antilinear $PT$ symmetry. As is characteristic of such systems, one can construct a probability conserving inner product that despite the linear in time growth is nonetheless time independent, and not only that, it leads to states with zero norm.  In addition, as is again characteristic of $PT$ symmetry,  these non-normalizable  states can be made normalizable by a continuation into a so-called Stokes wedge domain in the complex plane. In this domain one has a completely consistent quantum theory, one that lives alongside the standard normalizable energy eigenspectrum sector.  This thus not quite so  simple harmonic oscillator provides an explicit realization of  our general contention that antilinearity is more basic to quantum theory than Hermiticity.
\end{abstract}

\maketitle

\section{The various exact solutions to the simple harmonic oscillator}
 \label{S1}
 
 In this section we present the  various exact solutions to the simple harmonic oscillator that will be of interest to us in this paper. To begin we recall that the standard, non-relativistic, one-dimensional, quantum-mechanical, simple  harmonic oscillator is based on the Hamiltonian $\hat{H}=\hat{p}^2/2m+m\omega^2\hat{q}^2/2$ and quantization condition  $[\hat{q},\hat{p}]=i\hbar$. Its wave functions obey the Schr\"odinger equation
\begin{align}
i\hbar \frac{\partial \psi(q,t)}{\partial t}=\left[ -\frac{\hbar^2}{2m}\left(\frac{\partial}{\partial q}\right)^2+\frac{m\omega^2 q^2}{2}\right]\psi(q,t).
\label{1.1}
\end{align}
It has positive energy eigenvalues of the form $E_n=(n+1/2)\hbar \omega$,  with $n=0,1,2,3,...$, with eigenfunctions $\psi_n(q,t)=e^{-iE_nt/\hbar}H_n(q)e^{-q^2m\omega/2\hbar}$, where the $H_n(q)$ are Hermite polynomials. Each of these solutions is normalizable with $\int_{-\infty}^{\infty}dq \psi^*_n(q,t)\psi_n(q,t)$ being finite for each $n$.

Since the equation
\begin{align}
\left[ -\frac{\hbar^2}{2m}\left(\frac{\partial}{\partial q}\right)^2+\frac{m\omega^2 q^2}{2}\right]\psi_n(q,t)=E_n\psi_n(q,t)
\label{1.2}
\end{align}
is a second-order differential equation in $q$, for any given $E_n$ there must be two, thus degenerate, solutions. To find the second solution for the ground state we set $\psi(x,t)=e^{-iE_0t/\hbar}e^{-q^2m\omega/2\hbar}f(q)$. And with the prime denoting the derivative with respect to $q$, we find that $f(q)$ obeys 
\begin{align}
-\frac{\hbar^2}{2m}\left[f^{\prime\prime}-\frac{2qm\omega}{\hbar}f^{\prime}\right]=0, \qquad e^{q^2m\omega/\hbar}\frac{d}{dq}\left(e^{-q^2m\omega/\hbar}f^{\prime}\right)=0.
\label{1.3}
\end{align}
Other than the constant solution (viz. the one associated with the standard ground state) we obtain
\begin{align}
f^{\prime}(q)=e^{q^2m\omega/\hbar}, \qquad f(q)=\int^q dy e^{y^2m\omega/\hbar}=\left(\frac{m\omega}{\hbar}\right)^{1/2}\int^{(m\omega/\hbar)^{1/2}q}dz e^{z^2}.
\label{1.4}
\end{align}
We recognize $\int^{(m\omega/\hbar)^{1/2}q} dz e^{z^2}$ as the imaginary error function ${\rm erfi}((m\omega/\hbar)^{1/2}q)$.
To determine the asymptotic  behavior of the solution we set
\begin{align}
\frac{f(q)}{e^{q^2m\omega/\hbar}}=\frac{\int^q dx e^{x^2m\omega/\hbar}}{e^{q^2m\omega/\hbar}},
\label{1.5}
\end{align}
and using l'H\^opital's rule (ratio of derivatives for two divergent functions) obtain the asymptotic
\begin{align}
\frac{f(q)}{e^{q^2m\omega/\hbar}}\rightarrow \frac{e^{q^2m\omega/\hbar}}{2q (m\omega/\hbar)e^{q^2m\omega/\hbar}}= \frac{\hbar}{2qm\omega}.
\label{1.6}
\end{align}
Thus finally we have
\begin{align}
f(q)\rightarrow \frac{\hbar}{2qm\omega}e^{q^2m\omega/\hbar}, \qquad e^{-q^2m\omega/2\hbar}f(q)\rightarrow \frac{\hbar}{2qm\omega}e^{q^2m\omega/2\hbar}.
\label{1.7}
\end{align}
Even though this solution in (\ref{1.7}) diverges at large $q$, it has positive energy $E_0=\hbar\omega/2$, and should not be confused with the eigenfunction $e^{q^2m\omega/2\hbar}$ (viz. no additional $1/q$ factor) that holds for all $q$ and has negative energy $-\hbar\omega/2$. The solution we have in (\ref{1.7}) is strictly a positive energy solution.

To obtain the non-leading terms in our solution we recall the asymptotic power series expansion of ${\rm erfi}(z)$, viz. 
\begin{align}
\int^{z}dy e^{y^2}=\frac{e^{z^2}}{2z}\sum_{k=0}^{\infty} \frac{(2k-1)!!}{(2z^2)^k}=\frac{e^{z^2}}{2z}\left( 1+\frac{1}{2z^2}+\frac{3}{4z^4}+\frac{15}{8z^6}+...\right).
\label{1.8}
\end{align}
Thus for arbitrary $q$ we see that $\bar{\psi}_0(q)=e^{-q^2m\omega/2\hbar}f(q)$ is an odd function of $q$. Since the standard ground state eigenfunction $\psi_0=e^{-q^2m\omega/2\hbar}$ is an even function of $q$ it follows that $\int_{-\infty}^{\infty}dq \psi^*_0(q)\bar{\psi}_0(q)$ is zero. The two eigenfunctions are thus orthogonal, just as degenerate eigenfunctions should be.   Thus by construction we have shown that the simple harmonic oscillator has solutions that grow far more rapidly at spatial infinity than the oscillator potential itself, solutions that can thus escape from the oscillator altogether. We shall address this issue below, but note here that this would appear to be a general feature of potentials that grow as a power at infinity (such as  the linear one used in charmonium studies).

On having identified the $f(q)$ solution we can find yet another exact solution to the Schr\"odinger equation, one that is not an energy eigenstate of the oscillator at all, as would have to be the case since an energy eigenvalue  equation such as the second-order (\ref{1.2}) cannot have more than two energy eigenvalue solutions. To find this  solution we introduce 
\begin{align}
\hat{\psi}(q,t)=e^{-iE_0 t/\hbar}e^{-q^2m\omega/2\hbar}\left[ f(q)t+ig(q)\right],
\label{1.9}
\end{align}
and find that it obeys (\ref{1.1}) if 
\begin{align}
&-\frac{\hbar^2}{2m}\left[f^{\prime\prime}-\frac{2qm\omega}{\hbar}f^{\prime}\right]=0, \qquad -\frac{\hbar^2}{2m}\left[g^{\prime\prime}-\frac{2qm\omega}{\hbar}g^{\prime}\right]=\hbar f(q),
\nonumber\\
&g^{\prime}(q)=-\frac{2m}{\hbar}e^{q^2m\omega/\hbar}\int^qdy e^{-y^2m\omega/\hbar}f(y),\qquad g(q)=-\frac{2m}{\hbar}\int^qdye^{y^2m\omega/\hbar}\int^ydz e^{-z^2m\omega/\hbar}f(z).
\label{1.10}
\end{align}
Since $f(q)$ is an odd function of $q$, from (\ref{1.10}) it follows that $g(q)$ is an odd function of $q$ also, and with (\ref{1.7}) we find that $g(q)$ has  asymptotic limit
\begin{align}
g(q)\rightarrow -\frac{\hbar e^{q^2m\omega/\hbar} {\rm ln}q^2}{4m \omega^2 q},\qquad e^{-q^2m\omega/2\hbar}g(q)\rightarrow -\frac{\hbar e^{q^2m\omega/2\hbar} {\rm ln}q^2}{4m \omega^2q}.
\label{1.11}
\end {align}
Thus by explicit construction we have shown that the oscillator Schr\'odinger equation has a non-normalizable, non-stationary, solution that grows linearly in $t$. Since it is also an odd function of $q$ it is also orthogonal to the standard oscillator eigenfunction $\psi_0(q)=e^{-q^2m\omega/2\hbar}$. 

The $f,g$ type  pattern of solutions that we have found for the harmonic oscillator is quite general. We already noticed it for the square well, with it occuring  when bound states lie right at the top of the well. For the spherically symmetric three-dimensional  square well we find that $f\sim 1/r$ and $g\sim r$ \cite{Mannheim2025a}, and for the one-dimensional square well  $f\sim x$ and $g\sim x^3$.  In fact it even holds for a free theory. Specifically solutions to the free one-dimensional Schr\'odinger equation have energies of the form $E=\hbar^2k^2/2m$. They are twofold degenerate with eigenfunctions $\psi(k)=e^{ikx}$ and $\psi(-k)=e^{-ikx}$. However for $E=0$ there is only one solution, viz. $\psi(0)=1$. The missing solution is given as the limit $k\rightarrow 0$ of [$e^{ikx}-e^{-ikx}]/2ik=x$. It satisfies the Schr\"odinger equation but is not an eigenstate. Moreover, it is accompanied by yet another non-eigenstate solution, viz. $\psi(x,t)=tx-ix^3/6$. For special values of the energy we thus find a linear in time behavior. 

Having identified the various solutions of interest to us we turn now to a discussion of their significance. As we will see, things are not as straightforward as they seem.

\section{The $f(q)$ sector}
\label{S2}

To explore  the nature of the $f(q)$ solutions we introduce the standard creation and annihilation operators
\begin{align}
&\hat{q}=\left(\frac{\hbar}{2m\omega}\right)^{1/2}(a+a^{\dagger}), \qquad \hat{p}=i\left(\frac{\hbar m\omega}{2}\right)^{1/2}(a^{\dagger}-a),
\qquad [a,a^{\dagger}]=1,\qquad \hat{H}=\left(a^{\dagger}a+\frac{1}{2}\right)\hbar \omega,
\nonumber\\
&a=\left(\frac{m\omega}{2\hbar}\right)^{1/2}\hat{q}+i\left(\frac{1}{2\hbar m\omega}\right)^{1/2}\hat{p},\qquad a^{\dagger}=\left(\frac{m\omega}{2\hbar}\right)^{1/2}\hat{q}-i\left(\frac{1}{2\hbar m\omega}\right)^{1/2}\hat{p}.
\label{2.1}
\end{align}
The standard vacuum $\vert \Omega \rangle$ is defined as the state that obeys $a\vert \Omega \rangle=0$, $\hat{H}\vert \Omega \rangle =(\hbar\omega/2)\vert \Omega \rangle$. We introduce  a vacuum $\vert \bar{\Omega} \rangle$ for  the $f(r)$ sector. Since it is degenerate in energy with the standard $\vert \Omega \rangle$ it obeys 
\begin{align}
\hat{H}\vert \bar{\Omega} \rangle =\frac{\hbar\omega}{2}\vert \bar{\Omega} \rangle.
\label{2.2}
\end{align}
Since it is distinct from the standard $\vert \Omega \rangle$ it cannot be annihilated by $a$. Instead it must be annihilated by $a^{\dagger}a$, i.e., it must obey
\begin{align}
a^{\dagger}a\vert \bar{\Omega} \rangle =0.
\label{2.3}
\end{align}
It is thus (\ref{2.3}) that defines the $f(r)$ sector. (While the negative energy eigenstate $e^{q^2m\omega/\hbar}$ also grows exponentially it belongs to a sector built on a vacuum that  $a^{\dagger}$ annihilates -- viz. negative energy states propagating forward in time.)

To see the relation of the $\vert \bar{\Omega} \rangle$ vacuum to the standard $\vert \Omega \rangle$ vacuum, on recalling that  $\int_{-\infty}^{\infty}dq \psi^*_0(q)\bar{\psi}_0(q)=0$ we evaluate
\begin{align}
\langle \Omega  \vert \bar{\Omega} \rangle= \int _{-\infty}^{\infty}dq \langle \Omega\vert q\rangle \langle q   \vert \bar{\Omega} \rangle=\int _{-\infty}^{\infty}dq \psi_0^*(q)\bar{\psi}_0(q)=0.
\label{2.4}
\end{align}
The two vacua are thus orthogonal to each other, as is to be expected since they are associated with states that are degenerate in energy. However, there are 
actually two realizations  of (\ref{2.4}), either $\vert \bar{\Omega} \rangle$ and $\vert \Omega\rangle$ are two orthogonal vectors in the same Hilbert space, or they are vectors in totally different Hilbert spaces altogether and there is no meaning to the matrix element $\langle \Omega  \vert \bar{\Omega} \rangle$. We shall see that it is the latter option that is the case, and shall show this by proceeding with the former option and reaching a contradiction.

In analog to  the standard case the first excited state in the $f(r)$ sector is given by $a^{\dagger}\vert \bar{\Omega}\rangle$ since 
\begin{align}
\hat{H}a^{\dagger} \vert \bar{\Omega} \rangle= \left(a^{\dagger}a+\frac{1}{2}\right)\hbar \omega a^{\dagger}\vert \bar{\Omega} \rangle=\frac{1}{2}\hbar \omega a^{\dagger}\vert \bar{\Omega} \rangle+a^{\dagger}(1+a^{\dagger}a)\hbar \omega \vert \bar{\Omega} \rangle=\frac{3\hbar\omega}{2}a^{\dagger} \vert \bar{\Omega} \rangle.
\label{2.5}
\end{align}
We thus duplicate the standard treatment since it does not matter whether $a$ or $a^{\dagger} a$ annihilates the vacuum.

While this pattern duplicates the standard treatment for the general  $(a^{\dagger})^{ n} \vert \bar{\Omega} \rangle$ with it being an energy eigenstate with energy eigenvalue $(n+1/2)\hbar \omega$, there are some differences  for the eigenfunctions. To determine the first excited state eigenfunction $\langle q \vert a^{\dagger}\vert \bar{\Omega}\rangle=\bar{\psi}_1(q)$ we evaluate
\begin{align}
\bar{\psi}_1(q)&=\left[\left(\frac{m\omega}{2\hbar}\right)^{1/2}q-\left(\frac{\hbar}{2m\omega}\right)^{1/2}\frac{d}{d q}\right]\bar{\psi}_0(q)=\left[\left(\frac{m\omega}{2\hbar}\right)^{1/2}q-\left(\frac{\hbar}{2m\omega}\right)^{1/2}\frac{d}{d q}\right]\left[e^{-q^2m\omega/2\hbar}\int^q dx e^{x^2m\omega/\hbar}\right]
\nonumber\\
&=\left(\frac{2m\omega}{\hbar}\right)^{1/2}qe^{-q^2m\omega/2\hbar}\int^q dx e^{x^2m\omega/\hbar}-\left(\frac{\hbar}{2m\omega}\right)^{1/2}e^{q^2m\omega/2\hbar},
\label{2.6}
\end{align}
and one can readily check that it does indeed obey the Schr\"odinger equation with the requisite $3\hbar \omega/2$ eigenvalue.
Like  $\bar{\psi}_0(q)$, $\bar{\psi}_1(q)$ is also non-normalizable. However, unlike the standard $\psi_1(q)=\langle q\vert a^{\dagger}\vert \Omega\rangle=(2m\omega/\hbar)^{1/2}qe^{-q^2m\omega/2\hbar}$, $\bar{\psi}_1(q)$ is not given by the  Hermite polynomial $H_1(q)=q$ times $\bar{\psi}_0(q)$. Moreover, this is not even the case asymptotically since $\bar{\psi}_1(q)\rightarrow e^{q^2m\omega/2\hbar}/(2q^2(2m\omega/\hbar)^{1/2})\sim \bar{\psi}_0(q)/q$. However, since  $\bar{\psi}_1(q)$ is an even function of $q$ (it is constructed as an odd derivative of the even  $\bar{\psi}_0(q)$), and since the standard  $\psi_1(q)$ is an odd function  of $q$, it follows that $\int_{-\infty}^{\infty}dq \psi^*_1(q)\bar{\psi}_1(q)=0$, with the two first excited states being orthogonal to each other, just as should be the case for  states that are degenerate in energy. This same pattern then repeats for all the higher excited states.

Orthogonality between states with different energy eigenvalues would appear to be standard, since if we define left- and right-eigenvectors $\langle L\vert$ and $\vert R\rangle$ of $\hat{H}$ with respective eigenvalues $E_L$ and $E_R$ (we clarify our use of the left-right basis below), we obtain $\langle L\vert\hat{H}\vert R\rangle=E_L\langle L\vert R\rangle=E_R\langle L\vert R\rangle$, so that $\langle L\vert R\rangle$ is zero for any $E_L\neq E_R$.  Hence all the states built out of the $\vert \bar{\Omega}\rangle$ vacuum would appear to be orthogonal to all of the states built out of the standard $\vert \Omega\rangle$ vacuum. However, before we can draw such a conclusion we need to check whether the general $\langle L\vert R\rangle$ is in fact zero for states with $E_R\neq E_L$. To this end we choose a convenient pair, namely the first excited standard state in the $\vert\Omega\rangle$ sector  and the ground state in  the $\vert \bar{\Omega} \rangle$ sector, so chosen since both have even wave functions, so their overlap is not automatically zero. On dropping all irrelevant factors we obtain
\begin{align}
&\langle \psi_1\vert\bar{\psi}_0\rangle=\int_{-\infty}^{\infty}dq \psi^*_1(q)\psi_0(q)=\int_{-\infty}^{\infty}dq qe^{-q^2/2}e^{-q^2/2}\int^q dy e^{y^2}=-\frac{1}{2}\int_{-\infty}^{\infty}dq \frac{de^{-q^2}}{dq}\int^q dy e^{y^2}
\nonumber\\
&=-\frac{1}{2}\left[e^{-q^2}\int^q dy e^{y^2}\right]\bigg{|}^{\infty}_{-\infty}+ \frac{1}{2}\int_{-\infty}^{\infty}dq e^{-q^2} e^{q^2}=-\frac{1}{2}\left[e^{-q^2}\frac{e^{q^2}}{q}\right]\bigg{|}^{\infty}_{-\infty}+ \frac{1}{2}\int_{-\infty}^{\infty}dq =\frac{1}{2}\int_{-\infty}^{\infty}dq =\infty.
\label{2.7}
\end{align}
Thus rather than be zero the overlap is infinite. Thus while the relation $\langle L\vert\hat{H}\vert R\rangle=E_L\langle L\vert R\rangle=E_R\langle L\vert R\rangle$ would appear to be correct, it cannot be. Thus the  states in the $\vert\Omega\rangle$ sector and the states in the $\vert \bar{\Omega} \rangle$ sector cannot be connected by $\hat{H}$. Thus the two sets of states must exist in completely separate Hilbert spaces, with each one having its own Hamiltonian, even though they are identical in form. The sectors built on   $\vert\Omega\rangle$ and  $\vert \bar{\Omega} \rangle$ are thus associated  with inequivalent representations of the commutation algebra associated with the $\hat{q}$, $\hat{p}$ and $\hat{H}$ operators. We now construct the one associated with $\vert \bar{\Omega} \rangle$, and to do this we need to use $PT$ theory.

\section{Stokes wedges and $PT$ symmetry}
\label{S3}

The existence of non-normalizable wave functions and a procedure for dealing with them is encountered in the $PT$ symmetry program of  Bender and collaborators ($P$ is parity and $T$ is time reversal).  $PT$ symmetry has become a very active area of research as it provides an experimentally confirmed  generalization of conventional Hermitian quantum mechanics (see e.g. \cite{Bender1998,Bender1999,Mostafazadeh2002,Bender2002,Bender2007,Makris2008,Bender2008a,Bender2008b,Guo2009,Bender2010,Special2012,Theme2013,ElGanainy2018,
Bender2018,Mannheim2018a,Fring2021}, together with  more than 13,000 other peer-reviewed published papers to date). It is based on the realization that Hermiticity of a Hamiltonian is only sufficient to give real eigenvalues but not necessary, with the necessary condition  being that the Hamiltonian have an antilinear symmetry \cite{Bender2010,Mannheim2018a}. And  time reversal is such an antiliner operator. Similarly, Hermiticity of a Hamiltonian is only sufficient to give probability conservation, with the necessary condition  again being that the Hamiltonian have an antilinear symmetry \cite{Mannheim2018a}.  While Hamiltonians can have an antilinear symmetry (and because of the $CPT$ theorem where $C$ is charge conjugation many do), antilinear symmetry is more general, and in \cite{Mannheim2018a} it was shown to actually be the most general requirement that can lead to a consistent quantum mechanics.

As far as eigenvalues are concerned  we note that if we consider some general antilinear operator $\hat{A}$ that commutes with a Hamiltonian $\hat{H}$, and consider eigenvalues $E$ and eigenfunctions $e^{-iEt}\vert \phi\rangle$ of $\hat{H}$ that obey $\hat{H}\vert \phi \rangle=E\vert \phi\rangle$ we obtain
\begin{align}
\hat{H}\hat{A}\vert \phi\rangle=\hat{A}\hat{H}\vert \phi\rangle=\hat{A}E\vert \phi\rangle=E^*\hat{A}\vert \phi\rangle.
\label{3.1}
\end{align}
Thus for every eigenvalue $E$ with eigenvector $\vert \phi\rangle$ there is an eigenvalue $E^*$ with eigenvector $\hat{A}\vert \phi\rangle$. Thus as first noted by Wigner in his study of time reversal invariance \cite{Wigner1960}, antilinear symmetry implies that energy eigenvalues are either real or in complex conjugate pairs. Both of these possible realizations have been established (see e.g. \cite{Mannheim2025,Mannheim2025a} for some recent examples of the complex conjugate pair realization). As constructed, (\ref{3.1}) only shows that antilinear symmetry can lead to real eigenvalues, but in  \cite{Bender2010,Mannheim2018a}  it was shown to be necessary.

Even though we have shown that in the $\vert \bar{\Omega} \rangle$ sector the eigenvalues of the simple harmonic oscillator are all real, since the associated eigenfunctions are not normalizable, in this sector $\hat{H}$ is not self-adjoint.   Thus the reality of these eigenvalues cannot be attributed to Hermiticity. Rather it is due to a $PT$  symmetry that the simple harmonic oscillator can readily be seen to possess, even though it is not ordinarily considered. ($\hat{q}$ is $PT$ odd and $\hat{p}$ is $PT$ even, with  $\hat{H}=\hat{p}^2/2m+m\omega^2\hat{q}^2/2$ thus being  $PT$ even.) 

In regard to probability conservation, we note that while the matrix element $\langle\psi(t)\vert \psi(t)\rangle=\langle\psi(0)\vert e^{i\hat{H}^{\dagger}}(t)e^{-i\hat{H}t}\vert\psi(0)\rangle$ is not time independent if $\hat{H}$ is not Hermitian, that only means that we cannot use the standard Dirac inner product as the relevant norm for the theory. Instead we introduce a time-independent operator $\hat{V}$, and with $i\partial_t\vert \psi\rangle=\hat{H}\vert \psi\rangle$, $-i\partial_t\langle \psi \vert=\langle \psi \vert \hat{H}^{\dagger}$, we evaluate
\begin{align}
i\frac{\partial}{\partial t}\langle \psi(t)\vert \hat{V}\vert \psi(t)\rangle=\langle \psi(t)\vert (\hat{V}\hat{H}-\hat{H}^{\dagger}\hat{V})\vert \psi(t)\rangle.
\label{3.2}
\end{align}
Thus $\langle \psi(t)\vert \hat{V}\vert \psi(t)\rangle$ will be time independent and probability will be conserved 
 if there exists a $\hat{V}$ that obeys 
\begin{align}
\hat{V}\hat{H}=\hat{H}^{\dagger}\hat{V},\quad \hat{V}\hat{H }\hat{V}^{-1}=\hat{H}^{\dagger},
\label{3.3}
\end{align}
with the second condition requiring that $\hat{V}$ be invertible, something we take to be the case here. The $\hat{V}\hat{H}\hat{V}^{-1}=\hat{H}^{\dagger}$ condition is  known as pseudo-Hermiticity and implements probability conservation. Pseudo-Hermiticity was introduced by Dirac \cite{Dirac1942} and Pauli \cite{Pauli1943} in their study of indefinite metric quantum field theories (a recent discussion of which may be found in \cite{Mannheim2024}), and was connected to $PT$ symmetry in \cite{Mostafazadeh2002}.

The $\hat{V}\hat{H}\hat{V}^{-1}=\hat{H}^{\dagger}$ condition entails that the relation between $\hat{H}$ and $\hat{H}^{\dagger}$  is isospectral. Thus every eigenvalue of $\hat{H}$ is an eigenvalue of $\hat{H}^{\dagger}$. Consequently the eigenvalues of $\hat{H}$ are  either real or in complex conjugate pairs. But this is also the case with antilinear symmetry. Thus  antilinear symmetry and pseudo-Hermiticity are interchangeable. 

If we introduce right-eigenvectors $\vert R\rangle$ of $\hat{H}$ according to  $\hat{H}\vert R\rangle=E\vert R \rangle$, we have
\begin{align}
\langle R\vert \hat{H}^{\dagger}=E^*\langle R\vert=\langle R\vert \hat{V}\hat{H }\hat{V}^{-1},\qquad \langle R\vert \hat{V}\hat{H }=E^*\langle R\vert \hat{V}.
\label{3.4}
\end{align}
Consequently, we  can identify a left eigenvector $\langle L\vert=\langle R\vert \hat{V}$, and can write the inner product as $\langle R\vert \hat{V}\vert R\rangle=\langle L\vert R\rangle$. Thus in general we can  identify the left-right inner product as the most general probability-conserving inner product in the antilinear case, a form that could perhaps be  anticipated since a Hamiltonian cannot have any more eigenvectors than its left and right ones. Thus the way to generalize Hermitian quantum mechanics is to replace the dual space that is built out of the Hermitian conjugate bras $\langle R\vert$ of a given set of $\vert R\rangle$ kets by a dual space built on  $\langle R\vert \hat{V}=\langle L\vert$ bras instead. That we are able to do this is because the Schr\"odinger equation $\hat{H}\vert R\rangle=E\vert R \rangle$ only specifies the ket, leaving the bra undetermined. In general  the bra $\langle L\vert$ is not the Hermitian conjugate of $\vert R \rangle$, but is instead $\langle R\vert \hat{V}$, or equivalently the antilinear conjugate of $\vert R\rangle$. (Like antilinear conjugation Hermitian conjugation also involves complex conjugation, with antilinear conjugation thus being its natural generalization.) 

For a Hamiltonian that is not self-adjoint in a given coordinate domain the techniques of $PT$ theory enable one to find a different domain in which the Hamiltonian is self-adjoint (see e.g. \cite{Bender2007}). Thus for the $f(q)$ function given in (\ref{1.7}) $e^{-q^2m\omega/2\hbar}f(q)\rightarrow (\hbar/2qm\omega)e^{q^2m\omega/2\hbar}$ will converge if we replace $q$ by $\pm iq$, or more generally continue  $q$ into the so-called Stokes wedge given by the north and south quadrants of a letter $X$ drawn in the complex $q$ plane.  For operators we make the similarity transformation 
$e^{\pi\hat{p}\hat{q}/2}\hat{q}e^{-\pi\hat{p}\hat{q}/2}=-i\hat{q}=\hat{r}$, $e^{\pi\hat{p}\hat{q}/2}\hat{p}e^{-\pi\hat{p}\hat{q}/2}=i\hat{p}=\hat{s}$. Under this transformation we find that
\begin{align}
e^{\pi\hat{p}\hat{q}/2}\hat{H}e^{-\pi\hat{p}\hat{q}/2}=\tilde{H}=-\frac{\hat{s}^2}{2m}-\frac{m\omega^2\hat{r}^2}{2}, \qquad [\hat{r},\hat{s}]=i\hbar,\qquad 
i\hbar \frac{\partial \psi(r,t)}{\partial t}=-\left[ -\frac{\hbar^2}{2m}\left(\frac{\partial}{\partial r}\right)^2+\frac{m\omega^2 r^2}{2}\right]\psi(r,t).
\label{3.5}
\end{align}
Despite its overall minus sign,  wave functions such as $\hat{\psi}_0(r,t)=e^{-iE_0t}e^{r^2m\omega/2\hbar}\int^{\infty}_{r}dy e^{-y^2m\omega /\hbar}$ with well-behaved asymptotic limit $e^{-iE_0t}(\hbar/2rm\omega)e^{-r^2m\omega/2\hbar}$ ($\int_r^{\infty}dt e^{-t^2}\rightarrow e^{-r^2}/2r$) are still eigenstates of $\hat{H}$ with positive eigenvalues (as can readily be checked and as must be the case since one cannot change eigenvalues under a similarity transformation). For the general $\hat{\psi}(r,t)=e^{-iE_0t}e^{r^2m\omega/2\hbar}[f(r)t+ig(r)]$ we obtain
\begin{align}
&\frac{\hbar^2}{2m}\left[f^{\prime\prime}+\frac{2rm\omega}{\hbar}f^{\prime}\right]=0, \quad \frac{\hbar^2}{2m}\left[g^{\prime\prime}+\frac{2rm\omega}{\hbar}g^{\prime}\right]=\hbar f(r),
\quad f(r)=\int_r^{\infty}e^{-y^2m\omega/\hbar}dy=\left(\frac{m\omega}{\hbar}\right)^{1/2}\int^{\infty}_{(m\omega/\hbar)^{1/2}r}dz e^{-z^2},
\nonumber\\
&g^{\prime}(r)=\frac{2m}{\hbar}e^{-r^2m\omega/\hbar}\int^rdy e^{y^2m\omega/\hbar}f(y),\quad g(r)=\frac{2m}{\hbar}\int^rdye^{-y^2m\omega/\hbar}\int^ydz e^{z^2m\omega/\hbar}f(z),
\label{3.6}
\end{align}
with the needed convergent asymptotic behavior $f(r)\rightarrow \hbar e^{-r^2m\omega/\hbar}/2rm\omega$, $g(r)\rightarrow -\hbar e^{-r^2m\omega/\hbar}{\rm ln}r^2/4m\omega^2r$.
We thus see that the commutation algebra has two inequivalent representations, the standard one in  the east, west quadrants of the letter $X$ in the complex $q$ plane, quadrants that contain the real $q$ axis;  and the $PT$ one in its north, south quadrants, quadrants that contain the imaginary $q$ axis. Both realizations are valid and the energy eigenvalues are real and positive in both. Having now established the relevant $PT$ domain, we now show how to use the techniques of $PT$ theory to establish probability conservation for the modes given in (\ref{1.9}), modes  that grow linearly in time.

\section{The linear in time sector}
\label{S4}

For each eigenvalue there is an $f(r)$ type mode and an accompanying $f(r)t+ig(r)$ type mode, as exhibited n (\ref{1.4}) and (\ref{1.9}). However, only one of them is an eigenstate of the Hamiltonian. The eigenstates of the Hamiltonian thus do not span the full space of solutions to the theory and the Hamiltonian cannot be diagonalized. Such situations are known as Jordan-block situations, with Jordan having shown that any finite-dimensional matrix can either be diagonalized by a similarity transformation or brought to the Jordan canonical form in which all non-zero elements are equal to each other, with the non-zero ones filling out the main diagonal and one of its next to leading diagonals. The energies at which this occurs are known as exceptional points. Since there are two modes for a given $f(r)$ and $f(r)t+ig(r)$ pair, we can treat them as vectors in a two-dimensional space with Jordan-block matrix
\begin{align}
M=\begin{pmatrix}
1&1\\ 
0&1
\end{pmatrix}.
\label{4.1}
\end{align}
It has one left-eigenvector and one right-eigenvector with a zero norm overlap, viz.
\begin{align}
\begin{pmatrix}
1&1\\ 
0&1
\end{pmatrix}
\begin{pmatrix}
1\\ 
0
\end{pmatrix}=\begin{pmatrix}
1\\ 
0
\end{pmatrix},
\qquad \begin{pmatrix}
0&1\\ 
\end{pmatrix}\begin{pmatrix}
1&1\\ 
0&1
\end{pmatrix}=\begin{pmatrix}
0&1\\ 
\end{pmatrix},
\qquad \begin{pmatrix}
0&1\\ 
\end{pmatrix}\begin{pmatrix}
1\\ 
0
\end{pmatrix}=0.
\label{4.2}
\end{align}
With $M=\sigma_0+\sigma_+$, $[\sigma_0,\sigma_+]=0$, and $\sigma_+^2=0$, the time evolution associated with $M$ is given by
\begin{align}
&e^{-iMt}=e^{-i\sigma_0t}e^{-i\sigma_+t}=e^{-i\sigma_0t}[1-i\sigma_+t],\qquad e^{-iMt}\begin{pmatrix}
a\\ 
b
\end{pmatrix}=e^{-i\sigma_0 t}\begin{pmatrix}
a-itb\\ 
b
\end{pmatrix},
\label{4.3}
\end{align}
to thus naturally lead to the linear  time evolution behavior that we want.

For a generic $ft+ig$ type mode (one for the oscillator or square well for instance) we can construct explicit right- and left-eigenvector solutions to the Schr\"odinger equation associated with $M$, viz. 
\begin{align}
& i\frac{\partial}{\partial t}\begin{pmatrix}
e^{-it}(g-itf)\\ 
e^{-it}f
\end{pmatrix}=\begin{pmatrix}e^{-it}(g-itf+f)\\ 
e^{-it}f
\end{pmatrix}
=\begin{pmatrix}
1&1\\ 
0&1
\end{pmatrix}
\begin{pmatrix}
e^{-it}(g-itf\\ 
e^{-it}f
\end{pmatrix},
\label{4.4}
\end{align}
and
\begin{align}
& -i\frac{\partial}{\partial t}\begin{pmatrix}
e^{it}f^*,&
e^{it}(g^*+itf^*)\\
\end{pmatrix}=\begin{pmatrix}
e^{it}f^*,e^{it}(g^*+itf^*+f^*)& 
\\
\end{pmatrix}=\begin{pmatrix}e^{it}f^*,e^{it}(g^*+itf^*)&\\
\end{pmatrix}
\begin{pmatrix}
1&1\\ 
0&1
\end{pmatrix}.
\label{4.5}
\end{align}
The various  overlaps are given by
\begin{align}
&\begin{pmatrix}e^{it}f^*,e^{it}(g^*+itf^*)&\\
\end{pmatrix}\begin{pmatrix}
e^{-it}(g-itf)\\ 
e^{-it}f
\end{pmatrix}=f^*g+g^*f,\qquad \begin{pmatrix}
0,e^{it}\\ 
\end{pmatrix} \begin{pmatrix}
e^{-it}\\ 
0
\end{pmatrix}=0,
\nonumber\\
& \begin{pmatrix}e^{it}f^*,e^{it}(g^*+itf^*)&\\
\end{pmatrix}\begin{pmatrix}
e^{-it}\\ 
0
\end{pmatrix} =f^*,\quad \begin{pmatrix}0,e^{it}&\\
\end{pmatrix}\begin{pmatrix}
e^{-it}(g-itf\\ 
e^{-it}f
\end{pmatrix}=f.
\label{4.7}
\end{align}
These various overlaps are all time independent, as they must be since there does exist  a $V$ that effects $VMV^{-1}= M^{\dagger}$, viz. $V=\sigma_1$. Since $f(r)$ and $g(r)$ are bounded at large $r$ in the harmonic oscillator case (though not in the square well case where there are no Stokes wedges where wave functions are bounded), the subsequent integrations over $r$ are all finite. Thus in the north, south Stokes wedge in the complex $q$ plane we can construct an inner product in which $\langle L\vert=\langle R\vert \sigma_1$, with all $\langle L\vert R\rangle$ matrix elements being time independent, finite, and non-negative. The discussion repeats identically for the higher energy eigenvalues, with each pair of $f(r)$ type and $f(r)t+g(r)$ type modes forming its own two-dimensional Jordan-block matrix, and with all of these two-dimensional matrices being combined into one big block diagonal matrix. In this Stokes wedge we thus recognize each energy eigenvalue as being an exceptional point  \cite{footnote1}. 

To conclude we note that  with the simple harmonic oscillator having been studied since the very beginning of quantum mechanics, it is quite surprising not only to find a new set of harmonic oscillator solutions, but to find them to be of a Jordan-block $PT$ theory form rather than of a Hermitian theory form, with the simple harmonic oscillator not being quite so simple as it had been thought to be. Our study of the simple harmonic oscillator provides an explicit realization of  our general contention that antilinearity is more basic to quantum theory than Hermiticity.

\begin{acknowledgments}
The author wishes to thank  I. Sen for informing him of his own study on non-standard solutions to the harmonic oscillator \cite{Sen2024}, solutions that involve confluent hypergeometric functions, with these functions being related to the ${\rm erfi(q)}$.
\end{acknowledgments}

\end{document}